\documentclass[smallextended]{svjour3}       % onecolumn (second format)
\smartqed  % flush right qed marks, e.g. at end of proof
\usepackage{amsmath}
\usepackage{amssymb}
\usepackage[a4paper,margin=1.2in,footskip=0.25in]{geometry}
\usepackage{hyperref}
\usepackage{xcolor}
\usepackage{graphicx}
\newcommand{\orcid}[1]{\href{https://orcid.org/#1}{\includegraphics[width=10pt]{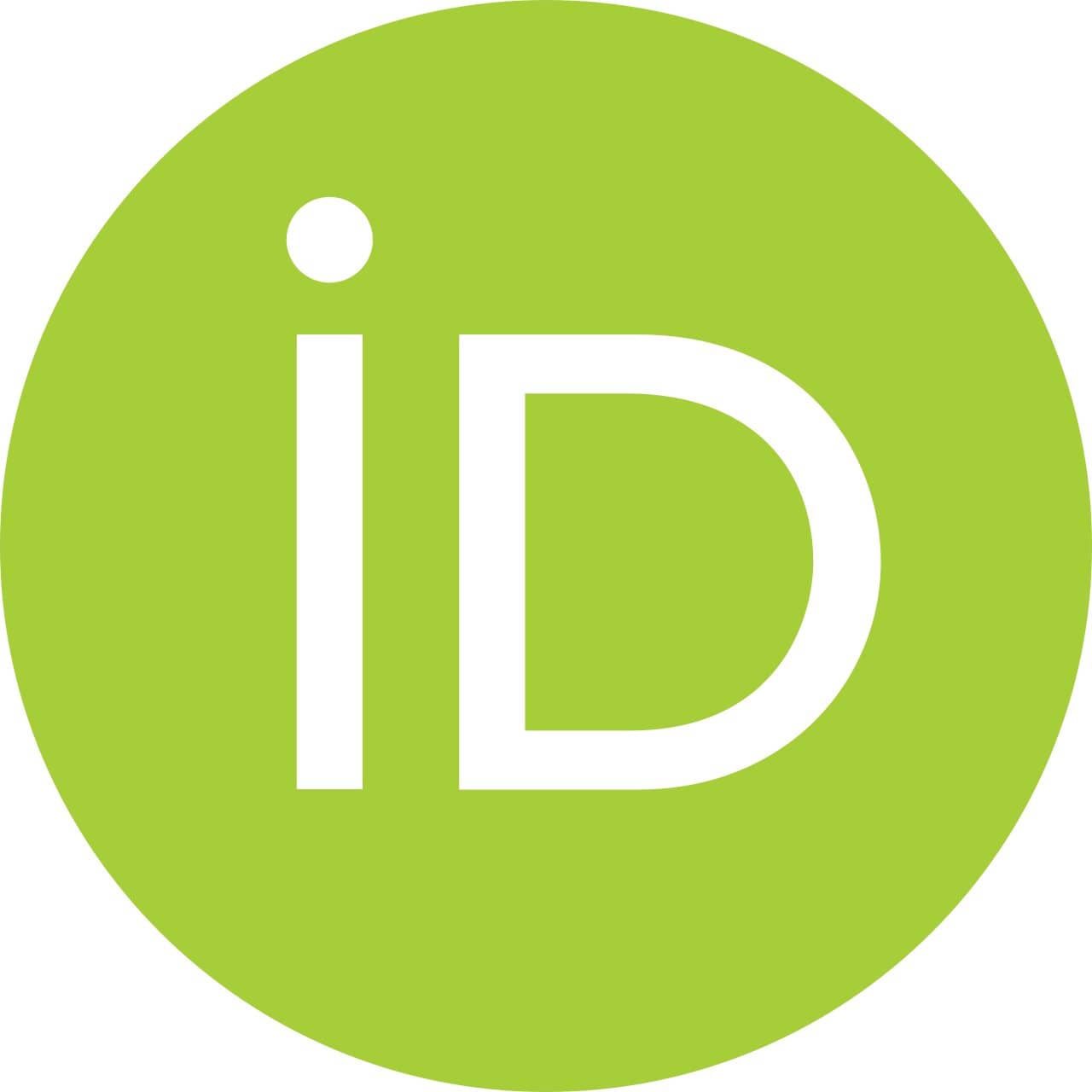}}}
\begin{document}
\title{Geometric quantification of multiparty entanglement through orthogonality of vectors%\thanks{Grants or other notes
}

%\titlerunning{Geometric quantification ...}        % if too long for running head

\author{Abhinash Kumar Roy$^{1}$  \orcid{0000-0001-7156-1989}, Nitish Kumar Chandra$^{1}$ \orcid{0000-0002-6572-1322}, S Nibedita Swain$^{1}$ \orcid{0000-0003-2326-2364}, Prasanta K. Panigrahi$^{1}$ \orcid{0000-0001-5812-0353}
}

\authorrunning{A. K. Roy, N. K. Chandra, S. N. Swain, P. K. Panigrahi} % if too long for running head

\institute{Prasanta K. Panigrahi (Corresponding Author)  \at
              \email{panigrahi.iiser@gmail.com}           %  \\
           \and
           Abhinash Kumar Roy  \at
              \email{akr16ms137@iiserkol.ac.in}
           \and
           Nitish Kumar Chandra  \at
              \email{nitishkrchandra@gmail.com} 
            \and
           S Nibedita Swain  \at
              \email{nibedita.iiser@gmail.com} 
            \and
          $^{1}\quad $\text{Indian Institute of Science Education and Research Kolkata, Mohanpur, West Bengal, 741246, India}
}

\date{Received: date / Accepted: date}

\maketitle

\begin{abstract}
The wedge product of vectors has been shown to yield the generalised entanglement measure I-concurrence, wherein the separability of the multiparty qubit system arises from the parallelism of vectors in the underlying Hilbert space of the subsystems. Here, we demonstrate the geometrical conditions of the post-measurement vectors which maximize the entanglement corresponding to the bi-partitions and can yield non-identical set of maximally entangled states. The Bell states for the two qubit case, GHZ and GHZ like states with superposition of four constituents for three qubits, naturally arise as the maximally entangled states. The geometric conditions for maximally entangled two qudit systems are derived, leading to the generalised Bell states, where the reduced density matrices are maximally mixed. We further show that the reduced density matrix for an arbitrary finite dimensional subsystem of a general qudit state can be constructed from the overlap of the post-measurement vectors. Using this approach, we discuss the trade-off between the local properties namely predictability and coherence with the global property, entanglement for the non-maximally entangled two qubit state. 
\keywords{Maximally entangled states \and Wedge product \and Complementarity}
\end{abstract}

\section{Introduction}
\label{intro}

Entanglement is one of the most distinctive features of quantum mechanics which manifests as nonlocal correlations in quantum systems \cite{schrodinger1935discussion,PhysRev.47.777,horodecki2009quantum,RevModPhys.84.1655}. %These nonlocal correlations make entanglement an important resource for quantum tasks . 
The maximally entangled Bell states have played a significant role in illustrating the nature of quantum correlations, as well as their usefulness in quantum tasks \cite{bell1964einstein,10.5555/1972505}. They have found applications in quantum cryptography \cite{ekert1991quantum}, quantum teleportation \cite{bennett1993teleporting}, 
super dense coding \cite{bennett1992communication}, to mention a few well known examples. The three particle Greenberger–Horne–Zeilinger (GHZ) state \cite{greenberger1989going}, its generalisations and W-states are known to show stronger non-locality and different entanglement properties \cite{agrawal2006perfect,Joo_2003}. These states have found use in quantum information splitting \cite{PhysRevA.59.1829,hsieh2010quantum,saha2012n}, quantum teleportation \cite{agrawal2006perfect,tsai2010teleportation,nandi2014quantum,adhikari2020probabilistic} and revealed subtle phase structure in atomic systems \cite{RevModPhys.84.1655,Parit:19}.
 Significantly, the implementation of quantum tasks has revealed the physical nature of the quantum non-local correlations \cite{vedral2014quantum}.
The entangled states of higher dimensions are under intense investigation as they provide advantage over qubit states and show better quantum correlations by strongly violating Bell's inequality \cite{erhard2020advances}. Entanglement has also thrown considerable light on correlations in spin chains and quantum many body systems \cite{PhysRevLett.91.207901,Das_2013,amico2008entanglement}.
Therefore, understanding of the physical nature of entanglement has attracted considerable attention in the current literature \cite{vedral2014quantum,mitra2015long,erhard2020advances}.

Significant amount of work has been carried out to quantify entanglement, which for the two particle case has been well understood \cite{horodecki2009quantum}. %However, entanglement in multipartite systems is yet to be accomplished.
While there exists several entanglement measures for pure and mixed two qubit systems such as, concurrence \cite{hill1997entanglement}, entanglement of formation \cite{wootters1998entanglement}, Schmidt number \cite{Sperling_2011} etc., there is no straightforward and unique way to characterize entanglement in multipartite setting. Among all these measures, concurrence is one of the most widely used, as it provides necessary and sufficient conditions for separability for a general pair of qubits. For arbitrary dimensions, the concurrence for multiparticle pure state was generalised by Rungta et al. and is known as "I-Concurrence"  \cite{rungta2001universal}. Mintert et al. carried out the general characterization of concurrence for multipartite quantum systems, showing that it is qualitatively different from other quantum correlations \cite{mintert2005concurrence}.

There have been various geometry based approaches to characterize and quantify entanglement \cite{OZAWA2000158,PhysRevA.95.032308,banerjee2019minimum,bhaskara2017generalized,banerjee2020quantifying}. Recently, a geometric perspective of I-concurrence has been given for multiparty entangled system of qudits, based on wedge product and Lagrange's identity, a generalisation of the Brahmagupta–Fibonacci identity \cite{doran2003geometric,brah}. This framework was further explored in \cite{banerjee2020quantifying}, where geometric representation of three way distributive entanglement, known as 3-tangle \cite{coffman2000distributed} in two dimensions is given, with a measure to quantify the distributive $n$-$\text{party}$ entanglement. In \cite{banerjee2019minimum}, the minimum distance between a set of bi-partite $n$-qudit density matrices with positive partial trace and maximally mixed states was calculated using a measure based on Euclidean distance between Hermitian matrices.

In this work, we use this wedge product formalism of I-concurrence to find geometrical configurations corresponding to the separable and maximally entangled states of finite dimensional multiparty systems. In this approach, entanglement across a bi-partition is given in terms of the wedge product of the post-measurement vectors. The problem of separability and entanglement is then reduced to geometry of area spanned by the post-measurement vectors.  Parallelism of post measurement vectors result in separability across the bi-partitions, whereas the orthogonality and equal norm of these vectors yield maximally entangled states. These constraints are employed to find maximally entangled states in case of two qubit and three qubit systems. For the two qubit case, one obtains the general expression of the maximally entangled states, which after local unitary transformations lead to the Bell states. In the three qubit case, starting with the Schmidt decomposed canonical forms \cite{acin2001three,acin2000}, we naturally obtain GHZ and GHZ like states \cite{yang2009quantum}, as maximally entangled states. For three qubit case, we also show that the parallelism across any two bipartitions is sufficient for tripartite separability. We then characterize and compare these states with the W-state using the polynomial invariant $3$-tangle {\cite{coffman2000distributed}}, defined in the wedge product formalism. The general conditions for maximally entangled states of a two-qudit system is established leading to the generalised Bell states. Obtaining the reduced density matrices corresponding to the subsystems in terms of overlap between the post-measurement vectors, we obtain the complementary relation between the global entanglement, $l_{1}$-norm coherence \cite{plenio}, and predictability for the two qubit systems.

The paper is organized as follows : In Sec. {\ref{prelim}} we review definitions of some basic quantities   In Sec. {\ref{Geometry}}, we illustrate the geometric approach to entanglement through two qubit systems and  describe the general configurations for minimizing maximizing the global entanglement of a finite dimensional multiparty system. We use these geometric constraints to find explicit conditions for separable and maximally entangled states of two and three qubit systems. Further, the geometric conditions for maximally entangled states in a general two-qudit system are obtained, which lead to the generalised Bell states. In Sec. \ref{vectors}, we investigate the non-maximally entangled two qubit pure state, and using geometric approach obtain the complementary relation between the global and local properties. We conclude in Sec. {\ref{sum}} with a summary of the work.

\section{Preliminaries}\label{prelim}
In this section, we provide some definitions which will be used throughout the paper.

\textit{Wedge Product}: Consider an $n$ dimensional space where $\left\{ e_{i}\right\}$, with, $i=1,2, ..n$, is an orthonormal basis. The linear superposition of two vectors $\Vec{a}=\sum_{i}a_{i}e_{i}$ and $\vec{b}=\sum_{j}b_{j}e_{j}$ forms a subspace, which naturally spans a complex plane containing the origin and these vectors. The wedge product of $\vec{a}$ and $\vec{b}$ is defined as:
\begin{equation}
\vec{a}\wedge\vec{b}=\sum_{i<j}\left(a_{i}b_{j}-a_{j}b_{i}\right)e_{i}\wedge e_{j} .
\end{equation}
The bivector $\vec{a}\wedge\vec{b}$ represents an oriented parallelogram in the plane, with adjacent sides as $\vec{a}$ and $\vec{b}$. Therefore, wedge product of two vectors naturally leads to a geometrical representation in terms of area in a plane. It is evident that the area spanned will be zero whenever the two vectors are parallel.   Maximizing the magnitude of the wedge product corresponds to maximizing the area of the parallelogram. Now, given a constraint on the perimeter of a parallelogram, or length of the vectors $\vec{a}$ and $\vec{b}$ (which in our case will be due to normalisation condition on states), the condition for maximal area is obtained through isoperimetric theorems \cite{benson1966euclidean}. One observes that, for some fixed length of vectors, area will be maximum when $\vec{a}$ and $\vec{b}$ are orthogonal, i.e., $\vec{a}\cdot\vec{b} = 0$. Furthermore, given a condition of the form, $|\vec{a}|^{n} + |\vec{b}|^{n}| = c$, where $c$ is a constant, the area is maximum when $|\vec{a}| = |\vec{b}|$. Therefore, maximum value of $|\vec{a}\wedge\vec{b}|$ corresponds to vectors $\vec{a}$ and $\vec{b}$ satisfying orthogonality and equal norm. This geometrically corresponds to the vectors $\vec{a}$ and $\vec{b}$, as the sides of a square in the underlying subspace.\\ A particularly useful identity concerning the wedge product is  Lagrange-Brahmagupta identity, which relates it to the norm and overlap of vectors \cite{stillwell2002elements},
\begin{equation}\label{LB}
    |\vec{a} \wedge \vec{b}|^{2} = |\vec{a}|^{2}|\vec{b}|^{2} - |\vec{a} \cdot \vec{b}|^{2},
\end{equation}
where, $|\cdot|$ stands for the norm.

\textit{Post-measurement vectors}: Consider a general $N$-qudit system with $S_{1},...,S_{N}$ representing individual qudits. The Hilbert space $\mathcal{H}_{S_{k}}$ of the qudit $S_{k}$ is $d_{k}$ dimensional with the standard basis represented by $\{|i_{k}\rangle\}$, where $i_{k}$ goes from $0$ to $d_{k}-1$. An arbitrary state representing the total system can be written as,
\begin{equation}
    |\Psi\rangle = \sum_{i_{1},..,i_{N}}C_{i_{1}i_{2}..i_{N}}|i_{1}\rangle\otimes|i_{2}\rangle\otimes\cdot\cdot\cdot\otimes|i_{N}\rangle,
\end{equation}
where, $C_{i_{1}i_{2}..i_{N}}$ are the complex coefficients subjected to normalization condition, and $|\Psi\rangle_{S_{1},...,S_{N}}\in \mathcal{H}_{S_{1}}\otimes\cdot\cdot\cdot\otimes\mathcal{H}_{S_{N}}$. Consider a bipartition $S_{1}....S_{j}|S_{j+1}...S_{N}$, then the post-measurement vectors corresponding to the subsystem $S_{1}....S_{j}$ is obtained as,
\begin{equation}
    \langle i_{1}\cdot\cdot\cdot i_{j}|\Psi\rangle = \sum_{i_{j+1},..,i_{N}}C_{i_{1}i_{2}..i_{N}}|i_{j+1}\rangle\otimes\cdot\cdot\cdot\otimes|i_{N}\rangle,
\end{equation}
where, $\langle i_{1}\cdot\cdot\cdot i_{j}| = \langle i_{1}|\otimes\cdot\cdot\cdot\otimes \langle i_{j}| $. As evident from the explicit expression of $\langle i_{1}\cdot\cdot\cdot i_{j}|\Psi\rangle$, these vectors lie in the Hilbert space $\mathcal{H}_{S_{j+1}}\otimes\cdot\cdot\otimes~\mathcal{H}_{S_{N}}$ of the subsystem $S_{j+1}...S_{N}$. Total number of post-measurement vector corresponding to $S_{1}....S_{j}$ will be $d_{1}\times\cdot\cdot\times d_{j}$. Similarly one may obtain the post-measurement vectors corresponding to other subsystems.

\textit{I-concurrence}: Consider a multiparty system described by a pure state $|\Phi\rangle$ . The I-concurrence corresponding to a partition $\mathcal{A}|\mathcal{B}$ is defined as:
\begin{equation}\label{concurr}
    C_{\mathcal{A}|\mathcal{B}}^{2} = 2(1-\operatorname{Tr}(\rho_{\mathcal{A}}^{2}))=  2(1-\operatorname{Tr}(\rho_{\mathcal{B}}^{2})),
\end{equation}
where, $\rho_{\mathcal{A}} = \operatorname{Tr}_{\mathcal{B}}(\rho)$ and $\rho_{\mathcal{B}} = \operatorname{Tr}_{\mathcal{A}}(\rho)$ are the reduced density matrices for subsystem $\mathcal{A}$ and $\mathcal{B}$ respectively with $\rho = |\Phi\rangle\langle\Phi|$.

\section{Geometric Conditions for separable and maximally entangled states} \label{Geometry}

Bhaskara and Panigrahi \cite{bhaskara2017generalized} have defined a global measure of entanglement as the sum of generalised concurrence corresponding to all the bi-partitions for a system using Lagrange's identity and showed that the concurrence of each bi-partition for a pure state can be quantified using wedge product of post-measurement vectors in the Hilbert space of the subsystems. This measure was shown to provide a faithful quantification of entanglement across any bi-partition for multipartite pure states. The wedge product formalism provides a geometric representation of concurrence, leading to the understanding of entanglement in terms of configuration of vectors in the Hilbert space of the subsystems. 

To illustrate, for the two qubit case, a state $|\Psi\rangle$ in general can be written as,
\begin{equation*}
    |\Psi\rangle_{AB} = |0\rangle\otimes|\psi_{1}\rangle + |1\rangle\otimes |\psi_{2}\rangle
\end{equation*}
where, $|\psi_{1}\rangle$ and $|\psi_{2}\rangle$ are two post-measurement vectors corresponding to subsystem $A$. In the language of geometric algebra, entanglement in the system $A|B$ is obtained in terms of wedge product of $|\psi_{1}\rangle$ and $|\psi_{2}\rangle$ as,
\begin{equation*}
    C_{A|B} = 2||\psi_{1}\rangle\wedge|\psi_{2}\rangle|
\end{equation*}
This definition facilitates a geometric description for entanglement, since $||\psi_{1}\rangle\wedge|\psi_{2}\rangle|$ is the magnitude of area spanned by the vectors $|\psi_{1}\rangle$ and $|\psi_{2}\rangle$ as shown in the figure below.
\begin{figure}[ht]
    \centering
    \includegraphics[width=0.5\textwidth]{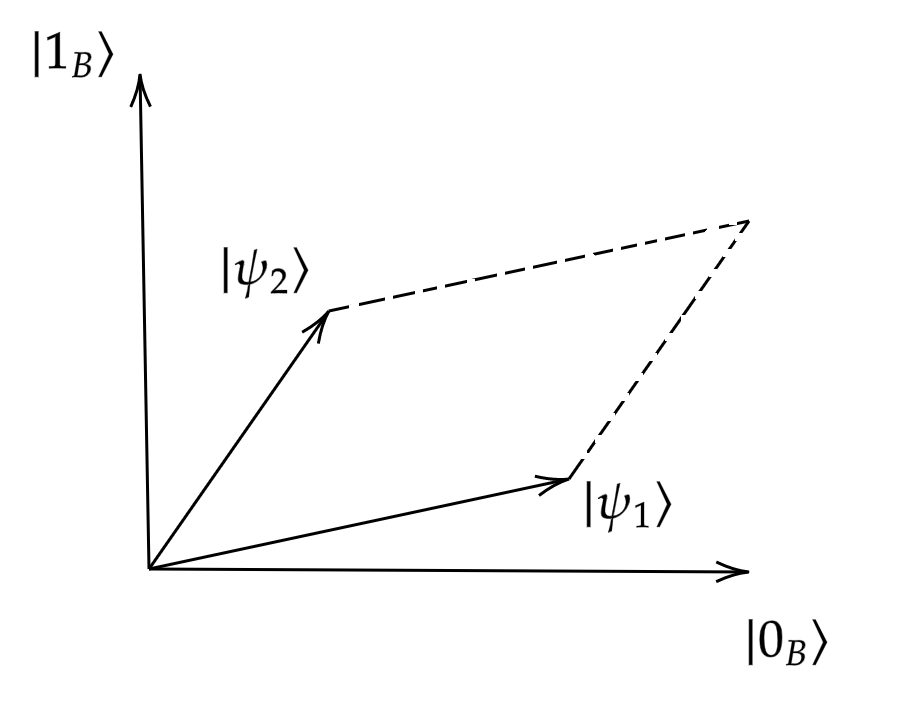}
    \caption{Area spanned by the post-measurement vectors $|\psi_{1}$ and $|\psi_{2}\rangle$  quantifies  entanglement in the bi-partition $A|B$}
    \label{area}
\end{figure}
Therefore, whenever the area spanned by the post-measurement vectors $|\psi_{1}\rangle$ and $|\psi_{2}\rangle$ is zero, the state will be separable. It corresponds to the scenario when $|\psi_{1}\rangle$ and $|\psi_{2}\rangle$ are parallel, i.e., $|\psi_{1}\rangle = \lambda|\psi_{2}\rangle$, where $\lambda \in \mathbf{C}$. In contrast, when the post-measurement vectors are non-parallel, the state will be entangled, and maximal entanglement corresponds to the configuration where area spanned is maximum. With a constraint on the perimeter (due to normalisation condition), as described above, through isoperimetric theorem in planar geometry one obtains that, the maximal area corresponds to the configuration where the vectors are sides of a square, i.e., orthogonal and equal. Therefore, the maximally entangled state will satisfy the conditions, $\langle\psi_{1}|\psi_{2}\rangle = 0$ and $||\psi_{1}\rangle| = ||\psi_{2}\rangle|$. In the following , we use the wedge product formalism to provide a geometrical perspective of entanglement in terms of parallelism of vectors for multipartite scenario.

Consider a multiparty system $\mathcal{S}$, described by a pure state $|\Psi\rangle$ and a bi-partition $\mathcal{A}|\mathcal{B}$, of this system, with Hilbert space corresponding to $\mathcal{A}$ and $\mathcal{B}$ given by $\mathcal{H}_{\mathcal{A}}$ and $\mathcal{H}_{\mathcal{B}}$ respectively. In terms of the the density matrix of the system, the I-concurrence, corresponding to the partition $\mathcal{A}|\mathcal{B}$ is given by Eq. (\ref{concurr}). Without loss of generality, one can take $d_{\mathcal{A}} = dim(\mathcal{H}_{\mathcal{A}}) \leq dim(\mathcal{H}_{\mathcal{B}})=d_{\mathcal{B}}$. If \{${|\phi_{i}\rangle}$\, $i$ = $1,2,.,d_{\mathcal{A}}$\} is an orthonormal basis of $\mathcal{H}_{\mathcal{A}}$, it follows that,
\begin{equation}
    |\Psi\rangle = \sum_{i = 0}^{d_{\mathcal{A}}}|\phi_{i}\rangle\otimes\langle\phi_{i}|\Psi\rangle.
\end{equation}

The set of post-measurement vectors $\{\langle\phi_{i}|\Psi\rangle\}$ contains $d_{\mathcal{A}}$ vectors, which are projections corresponding to each $|\phi_{i}\rangle \in \mathcal{H}_{\mathcal{A}}$, where the vectors $\langle\phi_{i}|\Psi\rangle\in\mathcal{H}_{\mathcal{B}}$. Concurrence of the bi-partition $\mathcal{A}|\mathcal{B}$, defined in terms of the wedge products of post-measurement vectors, is then \cite{bhaskara2017generalized} :
\begin{equation}
    C_{\mathcal{A}|\mathcal{B}}^{2} = 4\sum_{i<j}|\langle \phi_{i}|\Psi\rangle\wedge\langle \phi_{j}|\Psi\rangle|^{2},
\end{equation}
where $i$ and $j$ take values from $0$ to $d_{\mathcal{A}}$. The condition for separability across the bi-partition $\mathcal{A}|\mathcal{B}$ is $C_{\mathcal{A}|\mathcal{B}}^{2} = 0$. For vanishing concurrence, each of the wedge product, i.e., $\langle \phi_{i}|\Psi\rangle\wedge\langle \phi_{j}|\Psi\rangle~\forall~i,j$ must be zero, which in turn require the vectors $\langle \phi_{j}|\Psi\rangle$ to be parallel to each other, $\forall$ $i$ in the Hilbert space $\mathcal{H}_{\mathcal{B}}$, showing that the separability across bi-partitions can be viewed in terms of parallelism of vectors. For the maximally entangled state $|\Psi\rangle$, the I-concurrence $C_{\mathcal{A}|\mathcal{B}}$ must take maximum value for all the possible bi-partitions $\mathcal{A}|\mathcal{B}$. 

For a particular bi-partition  $\mathcal{A}|\mathcal{B}$, maximal $C_{\mathcal{A}|\mathcal{B}}$  corresponds to maximizing the magnitude of wedge product of post-measurement vectors, i.e, maximizing $|\langle \phi_{i}|\Psi\rangle\wedge\langle \phi_{j}|\Psi\rangle|$. As discussed in the previous section, it geometrically corresponds to maximizing the area spanned by the vectors $\langle \phi_{i}|\Psi\rangle$ and $\langle \phi_{j}|\Psi\rangle$. Using the result obtained through isoperimetric theorem, one obtains the geometrical condition for maximal $C_{\mathcal{A}|\mathcal{B}}$ as following,
\begin{equation}\label{ortho}
    \langle \phi_{i}|\Psi\rangle^{\dagger}\langle \phi_{j}|\Psi\rangle = 0, \quad \forall\quad i \neq j,
\end{equation}
\begin{equation}\label{equal}
    |\langle \phi_{i}|\Psi\rangle| = |\langle \phi_{j}|\Psi\rangle|, \quad \forall \quad i,j.
\end{equation}
   
Therefore, all the post measurement vectors are orthogonal to each other and have equal length in the Hilbert space $\mathcal{H}_{\mathcal{B}}$. Geometrically, it represents a $d_{\mathcal{A}}$ dimensional cube in the Hilbert space $\mathcal{H}_{\mathcal{B}}$. Now, a maximally entangled state $|\Psi\rangle$ must have maximal concurrence corresponding to all the possible bi-partitions, which occurs when the post measurement vectors for every bi-partition $\mathcal{A}|\mathcal{B}$ form a $d_{\mathcal{A}}$-cube in the Hilbert space $\mathcal{H}_{\mathcal{B}}$, where $d_{\mathcal{A}}\leq d_{\mathcal{B}}$.  

One thus obtains a set of constraints for separability and maximal entanglement, in terms of the relations between post-measurement vectors corresponding to bi-partitions of the system. The parallelism of these vectors corresponds to separability across the bi-partition, and orthogonality and equal norm of these vectors lead to maximal entanglement. 

We note in passing that the above geometrical conditions for maximal entanglement, when looked at in terms of reduced density matrix yield the absolute maximal entangled states \cite{Huber_2018}.  However, the Absolute Maximally Entangled state begins with the definition that the reduced density matrix corresponding to the smaller of the bi-partition is maximally mixed. In the usual approach, given the normalisation and positive semidefinite constraints on the density matrix, i.e., $\operatorname{Tr}(\rho) = 1$ and $\rho\geq 0$, maximizing the entropy of reduced density matrix leads to the condition that it must be proportional to identity. Whereas, in the present approach, this naturally comes from the geometry, namely maximizing the area spanned by the post-measurement vectors, wherein, the separability and entanglement across a bi-partition geometrically correspond to the post-measurement vectors being parallel and non-parallel, respectively. This approach is much simpler and transparent as compared to finding the reduced density matrix and checking its purity, particularly in the multiparty scenario. Geometrical configurations corresponding to maximal entanglement is derived by maximizing the area spanned without referring to a specific form of the reduced density matrix. Most importantly, the maximally mixed form of reduced density matrix for the maximally entangled case is not an assumption and, in fact, it naturally arises in our approach.

In the following subsections, we obtain the explicit conditions for separable and maximally entangled states for two qubit systems, three qubit systems and two particle qudit states.

\subsection{\textbf{The two qubit entanglement}}
For a two qubit system, the general state $|\psi\rangle$ in the computational basis is given by,
\begin{equation}
|\psi_{AB}\rangle=a|00\rangle+b|01\rangle+c|10\rangle+d|11\rangle,
\end{equation}
where,
$|i j\rangle=|i_{A}\rangle\otimes|j_{B}\rangle$ and $a,b,c,d\in\mathbb{C}$, satisfying the normalisation condition  $|a|^{2}+|b|^{2}+|c|^{2}+|d|^{2} = 1$.
Here, we have only bi-partition $A|B$ and, both $\mathcal{H}_{A}$ and $\mathcal{H}_{B}$ are two dimensional Hilbert spaces. The post measurement state corresponding to $\left|0\right\rangle$  and $|1\rangle$ for particle A are, 
$\left.\left\langle 0_{A}|\psi\right\rangle =a|0_{B}\rangle+b|1_{B}\right\rangle$  and $\left\langle 1_{A}|\psi\right\rangle =c\left|0_{B}\right\rangle +d|1_{B}\rangle$. These are vectors in the two dimensional complex Hilbert space $\mathcal{H_{B}}$.  The generalized concurrence measure in terms of wedge product  for $|\psi\rangle_{AB}$ is given by: 
\begin{equation}
C_{A|B} = 2\left|\left\langle 0_{A}\mid\psi\right\rangle \wedge\langle1_{A}|\psi\rangle\right|.
\end{equation}
%\begin{figure}[hbt!]
%    \centering
 %   \includegraphics[width=8cm]{p1.jpg}
 %   \caption{Bivector $\langle 0_{A}|\psi\rangle \wedge\langle1_{A}|\psi\rangle$ as an oriented parallelogram in the Hilbert space $\mathcal{H}_{B}$, with the vectors $\vec{OC}$ and $\vec{OA}$ representing $\langle 0_{A}|\psi\rangle$ and $\langle1_{A}|\psi\rangle$, respectively.}
 %   \label{fig:parallelogram}
%\end{figure}
 The separability condition is the parallelism of vectors $\langle 0_{A}|\psi_{AB}\rangle$ and $\langle 1_{A}|\psi_{AB}\rangle$ in the Hilbert space $\mathcal{H}_{B}$, which corresponds to $\frac{a}{c}= \frac{b}{d}$.  Maximizing $C_{A|B}$, geometrically corresponds to maximizing the area of the parallelogram formed by vectors $\vec{OC}=\langle 0_{A}|\psi_{AB}\rangle$ and $\vec{OA}=\langle 1_{A}|\psi_{AB}\rangle$, as two adjacent sides in $\mathcal{H}_{B}$. Using (\ref{ortho}) and (\ref{equal}) one obtains, $\Bar{a}=kd $ and $\Bar{b}=-kc$, where  $k=e^{i\theta}$.
 The general maximally entangled state takes the form
\begin{equation}
|\psi\rangle_{AB}=a\left|00\right\rangle +b\left|01\right\rangle -\Bar{b}e^{-i\theta}\left|10\right\rangle +\Bar{a}e^{-i\theta}\left|11\right\rangle.
\end{equation}
The reduced density matrices $\rho_{A} \text{ and } \rho_{B}$ are maximally mixed for the above state,
\begin{equation}
\rho_{A}=Tr_{B}(\rho) =\sum_{i= 0,1}\left\langle i_{B}|\rho|i_{B}\right\rangle = \frac{I}{2},
\end{equation}
 and the concurrence is found to be, $C_{A|B}=2||a|^{2}+|b|^{2}|=1$, as expected. Bell states are obtained by taking appropriate values of the parameters $a$, $b$ and $\theta$, which are also obtained by performing local unitary transformations. For instance, for $a,b\in \mathbb{R}$, the unitary transformation $I\otimes U|\psi\rangle_{AB}$ results in the Bell states up to a phase factor:
\begin{equation}
 |\psi_{\pm}\rangle = \frac{(\left|00\right\rangle \pm  \left|11\right\rangle)}{\sqrt{2}}\quad \text{for}\quad U = \sqrt{2} \begin{pmatrix} a & b\\ -b & a \end{pmatrix},
\end{equation}

\subsection{\textbf{The three qubit entanglement}}
\vspace{5pt}
A general state for a three qubit system is given by :
\begin{equation}
    \begin{aligned}
    |\psi_{ABC}\rangle = \lambda_{0}|000\rangle + \lambda_{1}|001\rangle +  \lambda_{2}|010\rangle + \lambda_{3}|011\rangle + \lambda_{4}|100\rangle\\ + \lambda_{5}|101\rangle + \lambda_{6}|110\rangle + \lambda_{7}|111\rangle
    \end{aligned}
\end{equation}
where, $\lambda_{i}\in \mathbb{C}$, $i = 0,1,...,7$  satisfy the normalisation condition, $\sum_{i = 0}^{7}|\lambda_{i}|^{2}=1$. For the state $|\psi_{ABC}\rangle$, we have three independent bi-partitions. Therefore, the global measure of entanglement $C$ is given by sum of concurrence corresponding to all the three bi-partitions \cite{bhaskara2017generalized} : 
\begin{equation}
    C = C_{A|BC} + C_{B|AC} + C_{C|AB}.
\end{equation}
%Here, $E$ is the sum of entanglement of all three bipartition. 
In the wedge product formalism, the concurrence is given by:
\begin{equation}
    C = 2\sum_{i = A,B,C}|\langle 0_{i}|\psi_{ABC}\rangle \wedge \langle 1_{i}|\psi_{ABC}\rangle|.
\end{equation}
As before, the separability condition across any bipartition is the parallelism of the post-measurement vectors, since the wedge product is zero in that case. Interestingly, for the three qubit case, separability across any two bipartition is sufficient for tripartite separability.  For instance, assuming $C_{A|BC} = 0 $ and $C_{B|AC} = 0$ leads to $  \langle 1_{A}|\Psi_{ABC}\rangle =  \alpha\langle0_{A}|\Psi_{ABC}\rangle $ and $\langle 1_{B}|\Psi_{ABC}\rangle =  \beta\langle0_{B}|\Psi_{ABC}\rangle$ respectively, where $\alpha, \beta \in \mathbb{C}$. The resulting state is then tripartite separable, 
\begin{equation}\label{trisep}
    |\Psi_{ABC}\rangle = (|0_{A}\rangle + \alpha|1_{A}\rangle) \otimes(|0_{B}\rangle + \beta|1_{B}\rangle) \otimes |\psi_{C}\rangle
\end{equation}
where, $|\psi_{C}\rangle = \langle0_{A}0_{B}|\Psi\rangle$. For Eq. (\ref{trisep}),  $\langle 1_{c}|\Psi_{ABC}\rangle =  \gamma\langle0_{c}|\Psi_{ABC}\rangle$, where $\gamma\in C$, therefore $C_{AB|C} = 0$. Hence, vanishing of the concurrence across any two bi-partition implies vanishing of concurrence across the third bi-partition.  

%$E_{A|BC} = E_{B|AC} = 0 \Rightarrow E_{C|AB} = 0 $.

 For a bi-separable state three qubit state, concurrence across one of the bi-partition is zero and non-zero across other bi-partitions. For instance, for a bi-separable state with separability across the bi-partition $A|BC$, the post-measurement vectors  $\langle 0_{A}|\psi_{ABC}\rangle = \lambda_{0}|00\rangle + \lambda_{1}|01\rangle +  \lambda_{2}|10\rangle + \lambda_{3}|11\rangle$ and$ \langle 1_{A}|\psi_{ABC}\rangle = \lambda_{4}|00\rangle + \lambda_{5}|01\rangle + \lambda_{6}|10\rangle + \lambda_{7}|11\rangle$ must be parallel, i.e.,
\begin{equation}
    \frac{\lambda_{0}}{\lambda_{4}} = \frac{\lambda_{1}}{\lambda_{5}} = \frac{\lambda_{2}}{\lambda_{6}} = \frac{\lambda_{3}}{\lambda_{7}}.
\end{equation}
It can be seen that global entanglement for bi-separable states is upper bounded by two. Therefore, if the global entanglement $C$ is greater than two, the state cannot be separated across any bi-partition, i.e., genuine tripartite entangled.\\
To obtain the maximally entangled states, one needs to maximize $|\langle 0_{i} | \psi_{ABC}\rangle \wedge\langle 1_{i} | \psi_{ABC}\rangle|$  for all $i=A,B,C$. The conditions for maximally entangled states as discussed previously are : $\langle 0_{i}| \psi_{ABC}\rangle$ must be orthogonal to $\langle 1_{i}|\psi_{ABC}\rangle$  and $|\langle 0_{i}|\psi_{ABC}\rangle| = |\langle 1_{i}|\psi_{ABC}\rangle|$ for each $i = A, B, C$, leading to the following conditions,
\begin{equation}
    \begin{array}{cc}
        \lambda_{0}\Bar{\lambda_{4}} + \lambda_{1}\Bar{\lambda_{5}} + \lambda_{2}\Bar{\lambda_{6}} + \lambda_{3}\Bar{\lambda_{7}} = 0,&  \\
         \lambda_{0}\Bar{\lambda_{2}} + \lambda_{1}\Bar{\lambda_{3}} + \lambda_{4}\Bar{\lambda_{6}} + \lambda_{5}\Bar{\lambda_{7}} = 0,&  \\
         \lambda_{0}\Bar{\lambda_{1}} + \lambda_{2}\Bar{\lambda_{3}} + \lambda_{4}\Bar{\lambda_{5}} + \lambda_{6}\Bar{\lambda_{7}} = 0;&
    \end{array}
\end{equation}
and,
\begin{equation}
\
    \begin{array}{cc}
      |\lambda_{0}|^{2} + |\lambda_{1}|^{2} + |\lambda_{2}|^{2} + |\lambda_{3}|^{2}  = |\lambda_{4}|^{2} + |\lambda_{5}|^{2} + |\lambda_{6}|^{2} + |\lambda_{7}|^{2}, &  \\
        |\lambda_{0}|^{2} + |\lambda_{1}|^{2} + |\lambda_{4}|^{2} + |\lambda_{5}|^{2}  = |\lambda_{2}|^{2} + |\lambda_{3}|^{2} + |\lambda_{6}|^{2} + |\lambda_{7}|^{2},  & \\
         |\lambda_{0}|^{2} + |\lambda_{2}|^{2} + |\lambda_{4}|^{2} + |\lambda_{6}|^{2}  = |\lambda_{1}|^{2} + |\lambda_{3}|^{2} + |\lambda_{5}|^{2} + |\lambda_{7}|^{2}. &
    \end{array}
\end{equation}
 Acín et al. have shown that any pure three qubit can be reduced to a canonical form, having five amplitudes and one relative phase using the generalised Schmidt decomposition \cite{acin2001three,acin2000}. Any generic three qubit state can be reduced to several classes of canonical forms using five local base product states (LBPS), while preserving nonlocal properties of the state. Owing to different degrees of orthogonality, there exists three inequivalent classes of five LBPS.
We begin with the general state built using the canonical forms from these LBPS classes.% Once a maximally entangled state is obtained, we perform local unitary (LU) transformations to find other states that form a complete basis in the Hilbert space of three qubit system, as LU operations do not affect entanglement of the state \cite{kraus2010local}.

One such canonical form built from a symmetric LBPS class, with qubits A, B and C is given by :
\begin{equation}
|\psi\rangle_{ABC}= a e^{i\theta}\left|000\right\rangle +b\left|001\right\rangle +c\left|010\right\rangle +d\left|100\right\rangle +e\left|111\right\rangle, 
\end{equation}
where $a,b,c,d,e$ are positive numbers and $0\leq\theta\leq\pi$.

To obtain maximally entangled states, we use the othogonality and equal norm constraints for all the three bi-partitions $A|BC, B|AC$ and $C|AB$. The orthogonality conditions on the post measurement vector yields, $ab = ac = ad  = 0$ and the equal norm conditions after some calculation leads to, $|b| = |c| = |d| \quad \text{and} \quad |a|^{2} + |d|^{2} = |e|^{2}.$
Two set of equations satisfy the above conditions.
%\begin{enumerate}
%\item 
For $a \neq0$, we get the maximally entangled state GHZ state,
  \begin{equation}
    |\psi\rangle=\frac{1}{\sqrt{2}}(|000\rangle+|111\rangle).
\end{equation}
%This maximally entangled state is the well known GHZ State.
  %\item
  For $a=0$,
%\begin{equation}
%    |\lambda_{1}|=|\lambda_{2}|=|\lambda_{3}|=|\lambda_{4}| = \frac{1}{2}
%\end{equation}
%Since the $k_{i}$'s are positive numbers, 
the maximally entangled state obtained is :
\begin{equation}
    |\xi_{1}\rangle= \dfrac{1}{2}(|001\rangle+|010\rangle+|100\rangle+|111\rangle)
\end{equation}
%\end{enumerate}
This state is known in the literature as GHZ like state . It has maximal entanglement measure and satisfies the condition $\rho_{A} = \rho_{B} = \rho_{C} = \frac{I}{2}$. These states can be derived alternatively using the algebraic properties of density matrix \cite{doi:10.1142/S0129055X14500044}.

Using the entanglement measure of \cite{bhaskara2017generalized}, we have been able to find states corresponding to the maximally entangled states as the geometry corresponding to the extrema (separable and maximally entangled states) are well defined. Given a non-extremal global entanglement, a natural configuration cannot be assigned to the post-measurement vectors since various combination of vector norms and angles between the post-measurement vectors can span same area. Therefore, one cannot find states that do not correspond to maximum value of the measure, the highly entangled W-state is one such example,
\begin{equation}
    |W\rangle = \frac{1}{\sqrt{3}}(|100\rangle + |010\rangle + |001\rangle).
\end{equation}

W-state satisfies all the orthogonality criteria for the maximally entangled state, but does not satisfy the equality of sides criterion, and hence its configuration does not correspond to that of maximal entanglement. The length of the post measurement vectors are unequal and are in the ratio of $\sqrt{2}:1$. For states which do not belong to the extremum ones, we cannot ascribe the geometry of the state without knowing the state a prori. The global entanglement for W state is not maximal and has value $2\sqrt{2}$ as compared to $3$ for GHZ and GHZ like states. %It is worth noting that in terms of persistence of entanglement, the GHZ like states are more like W- state, as both show robustness of entanglement under loss of any single qubit. This interesting property of GHZ like states differentiates it from GHZ states. 

In order to characterise the genuine three party distributive entanglement in three qubit pure states $|\psi_{ABC}\rangle$, one can use the 3-tangle, which is a polynomial invariant quantity under permutation of qubits, follows from the Coffman–Kundu–Wooter's inequality $C_{A \mid B C}^{2}\geq C_{A \mid B}^{2}+C_{A \mid C}^{2}$ \cite{coffman2000distributed}. Here $C_{A \mid B C}^{2}$ corresponds to the squared concurrence corresponding to the bi-partition $A| BC$, and $C_{A \mid B(C)}^{2} $ is the square concurrence between A and B(C). If the qubits $A$, $B$ and $C$ are entangled, the residual entanglement or 3-tangle is given by :
\begin{equation}
\tau=C_{A \mid B C}^{2}-C_{A \mid B}^{2}-C_{A \mid C}^{2}
\end{equation} 
 In \cite{banerjee2020quantifying}, this was defined in terms of wedge product \cite{banerjee2020quantifying}, where the post-measurement vectors considered were obtained corresponding to the subsystem with higher dimensional Hilbert space, say for a bi-partition $A|BC$, measurement was done on the bi-partition $BC$, this resulted into four post-measurement vectors in a two dimensional planar Hilbert space, which lead to geometrical manifestation of 3-tangle in terms of area inequality. Since in our analysis, we are considering the smaller of the bipartitions, the tangle can be alternatively defined with measurements corresponding to smaller of the two bi-partitions, where, the 3-tangle takes the form:
\begin{equation}
\begin{aligned} 
\tau = 4|\langle 0_{A}|\psi_{ABC}\rangle\wedge\langle 1_{A}|\psi_{ABC}\rangle|^{2}-4|\sum_{i = 0,1}\langle 0_{A}|\psi_{AB}\rangle_{i}\wedge\langle 1_{A}|\psi_{AB}\rangle_{i}|^{2} \\
-4|\sum_{i = 0,1}\langle 0_{A}|\psi_{AC}\rangle_{i}\wedge\langle 1_{A}|\psi_{AC}\rangle_{i}|^{2},
\end{aligned}
\end{equation}
where, $|\psi_{AC}\rangle_{i} = \langle i_{B}|\psi_{ABC}\rangle   \quad \text{and} \quad  |\psi_{AB}\rangle_{i} = \langle i_{C}|\psi_{ABC}\rangle$ are the post measurement vectors corresponding to $B$ and $C$ respectively. For the GHZ and GHZ like states, the 3-tangle obtained is one and vanishes for the W-state.   %These states belong to two inequivalent classes under stochastic local operations and classical communications (SLOCC) \cite{PhysRevA.62.062314}.

\subsection{\textbf{The two qudit entanglement}} \label{qudit}
Consider the general form of two particle qudit state,
\begin{equation}
    |\psi_{AB}\rangle = \sum_{i,j}a_{ij}|i_{A}\rangle\otimes|j_{B}\rangle,
\end{equation}
where $i, j = 0,...,d-1$ for a $d$ -level computational basis. 
%In terms of $d$ post-measurement state vectors  $\langle i_{A}|\psi_{AB}\rangle$ in the $d$-dimensional Hilbert space $\mathcal{H}_{B}$, measure of entanglement of A with B in terms of wedge product is defined as:
%\begin{equation}
%    E^{2} = 4\sum_{i<j}|\langle i_{A}|\psi_{AB}\rangle\wedge\langle j_{A}|\psi_{AB}\rangle|^{2}
%\end{equation}
%As has been done earlier, for maximally entangled state, one needs to maximize the magnitude of wedge product between all the post-measurement state vectors; this leads to the following conditions:
%\begin{gather}
 %   \langle i_{A}|\psi_{AB}\rangle^{\dagger}\langle j_{A}|\psi_{AB}\rangle =  0 \quad i \neq j,\\
%    |\langle i_{A}|\psi_{AB}\rangle| = |\langle j_{A}|\psi_{AB}\rangle| \quad \forall \quad i,j.
%\end{gather}
From the discussion in the beginning of the section, the maximal entanglement will correspond to the configuration when vectors $\langle i_{A}|\psi_{AB}\rangle$ are the sides of a d-cube in the Hilbert space $\mathcal{H}_{B}$. Interestingly, for the maximally entangled case, as the vectors $\langle i_{A}|\psi_{AB}\rangle$ are the sides of a d-cube in $\mathcal{H}_{B}$, one can consider these vectors as a set of new computational basis and write $|\psi_{AB}\rangle$ in a Schmidt decomposed form. Taking $|i^{'}_{B}\rangle = \sqrt{d}\langle i_{A}|\psi_{AB}\rangle$ as the new orthonormal basis, we get the state vector in the following form,
$|\psi_{AB}\rangle = \sum_{i}a^{'}_{i}|i_{A}\rangle\otimes|i_{B}\rangle$, with $a^{'}_{i} = \frac{1}{\sqrt{d}}$. Hence, for a two-qudit system, the maximally entangled state takes the form,
\begin{equation}
    |\psi_{AB}\rangle = \frac{1}{\sqrt{d}}\sum_{i = 0}^{d-1}|ii\rangle.
\end{equation} 
This is a generalised form of the Bell states, as is evident from the fact that the partial trace corresponding to subsystem A or B, leads to $I/d$ where, $I$ is the identity matrix. As an example, for a two qutrit state, the maximally entangled state is obtained as:
\begin{equation}
    |\psi_{++}\rangle = \frac{1}{\sqrt{3}}(|00\rangle + |11\rangle + |22\rangle).
\end{equation}
The complete orthogonal basis of maximally entangled states can be found using local unitary transformations $U_{1}\otimes U_{2}$ on $|\psi_{++}\rangle$.

For an $n$-qudit system, total number of bi-partitions are $2^{n-1}-1$. A maximally entangled state must satisfy the constraints of orthogonality and equal norm of vectors for each of these bi-partitions. However, it is not guaranteed that such a state necessarily exists. In fact, maximally entangled state does not exist for a four qubit and more than six qubit systems \cite{Huber_2018}. It is seen that the GHZ state for four qubits:
\begin{equation}
    |GHZ\rangle_{4} = \frac{1}{\sqrt{2}}(|0000\rangle + |1111\rangle)
\end{equation}
does not satisfy the equal norm criterion for the post measurement vectors for the bi-partitions, where each subsystem has two qubits and hence, does not achieve the maximum value for global entanglement. This scenario also occurs for the $|GHZ\rangle_{n}= \frac{1}{\sqrt{2}}(|00...0_{n}\rangle + |11...1_{n}\rangle)$,
describing the n-qubit system with $n \geq 4$.
\section{Non-maximally entangled states and complementarity}\label{vectors}
Consider a two qubit system with three non-zero coefficient in the computational basis, for instance,
\begin{equation}
    |\psi\rangle = a|00\rangle + b|01\rangle + c|10\rangle,
\end{equation}
with, $a,b,c \in \textbf{C}$ satisfying the normalisation condition $|a|^{2} + |b|^{2} + |c|^{2} = 1$ and $a, b, c \neq0$. 
The post-measurement vectors corresponding to the subsystem $A$ are, $|\phi_{1}\rangle = \langle 0_{A}|\phi\rangle = a|0\rangle_{B} + b|1\rangle_{B}$ and $|\phi_{2}\rangle = \langle 1_{A}|\phi\rangle = c|0\rangle_{B}$. Therefore, one observes that for three non-zero coefficients in a general two qubit system in its computational basis, the state is always non-separable and also non-maximally entangled, since for such a scenario neither of the condition, parallelisability or orthogonality holds. At this point, one might ask, what makes these states distinct from the maximally entangled ones such as, the Bell states and, whether the deficiency in the global entanglement compensated through other local characteristics of the system. The answer to the latter turns out to be affirmative and, it is observed that the the local properties of coherence and predictability compensate the deficiency in the global entanglement from maximal, for the non-maximally entangled systems and obeys a triality relation \cite{jacob2010}, which we investigate in the following.
 
Firstly, we show that the reduced density matrix of the subsystems can be written in terms of inner product of post-measurement vectors. Suppose, $|\Psi\rangle$ describes a multiparty system $\mathcal{S}$ with the Hilbert space $\mathcal{H}$. We consider an arbitrary partition $\mathcal{A}|\mathcal{B}$ of $\mathcal{S}$, such that, $\mathcal{H} = \mathcal{H}_{\mathcal{A}}\otimes\mathcal{H}_{\mathcal{B}}$, with $d_{\mathcal{A}} = dim(\mathcal{H}_{\mathcal{A}})$ and $d_{\mathcal{B}} = dim(\mathcal{H}_{\mathcal{B}})$, where subsystem $\mathcal{A}$ and $\mathcal{B}$ can be any finite collection of qudits. Suppose, $(|\phi_{i}\rangle_{\mathcal{A}})_{i = 1}^{d_{\mathcal{A}}}$ form an orthonormal basis of the Hilbert space $\mathcal{H}_{\mathcal{A}}$, then the post measurement vectors corresponding to subsystem $\mathcal{A}$ are obtained as $|\psi_i\rangle_{\mathcal{B}} = \langle\phi_i|\Psi\rangle$ where $i= 1,..,d_{\mathcal{A}}$. The post-measurement vectors $|\psi_{i}\rangle_{\mathcal{B}}$ lie in the Hilbert space $\mathcal{H}_{\mathcal{B}}$ . The state vector of the system is :
\begin{equation}
    |\Psi\rangle = \sum_{i = 1}^{d_{\mathcal{A}}}|\phi_{i}\rangle_{\mathcal{A}}\otimes|\psi_{i}\rangle_{\mathcal{B}}.
\end{equation}
Reduced density matrix of subsystem $\mathcal{A}$ is given by $\rho^{\mathcal{A}} = \operatorname{Tr}_{\mathcal{B}}\rho$, where $\rho = |\Psi\rangle\langle\Psi|$ and,
\begin{align}
    \rho^{\mathcal{A}}_{ij} = \langle\phi_{i}|\rho^{\mathcal{A}}|\phi_{j}\rangle
                    =\sum_{m=1}^{d_{\mathcal{B}}}\langle\xi_{m}|\psi_{i}\rangle\langle\psi_{j}|\xi_{m}\rangle,
\end{align} 
where  $(|\xi_{i}\rangle_{\mathcal{B}})_{i = 1}^{d_{\mathcal{B}}}$ is an orthonormal basis of $\mathcal{H}_{\mathcal{B}}$. Using completeness of the basis, it follows that the reduced density matrix $\rho^{\mathcal{A}}$ in terms of overlap of post-measurement vectors is given by  $\rho^{\mathcal{A}}_{ij} = \langle\psi_{j}|\psi_{i}\rangle$. Therefore the diagonal elements corresponds to the length of the post-measurement vectors, equivalently, the probabilities of the possible outcomes.

%For the extreme cases (maximal coherence and zero coherence), we can see that the conditions are similar but opposite to that of maximally entangled states and separable states. When all the post measurement vectors, corresponding to the subsystem $\mathcal{A}$ are orthogonal and equal, $P_{\mathcal{A}}$ vanishes, and when they are parallel, $P_{\mathcal{A}}$ leads to maximum value one. Hence, intrinsic degree of coherence can be used to infer about separability and entanglement of the  state describing any two qudit system, where separable states and maximally entangled state will correspond to maximum and maximum intrinsic coherence for the subsystem respectively.

Consider a two qubit pure state $|\Psi\rangle$, with the post-measurement vectors for subsystem $\mathcal{A}$ corresponding to $|0_{A}\rangle$ and $|1_{A}\rangle$ denoted by, $|\psi_{1}\rangle$ and $|\psi_{2}\rangle$ respectively. As discussed in the previous section, global entanglement between subsystems $A$ and $B$ is given by,
\begin{equation}
    C_{A|B}^{2}  = 4||\psi_{1}\rangle \wedge |\psi_{2}\rangle|^{2}
\end{equation}
A quantitative measure of quantum coherence is $l_{1}$-norm, which satifies the non-negativity, monotonicity and convexity criteria \cite{plenio}, is defined in terms of off-diagonal elements of the density matrix. For the subsystem $A$, the coherence is obtained as,
\begin{equation}
    \mathcal{C}_{A} = \sum_{i \neq j} |\rho^{A}_{ij}| = 2|\langle\psi_{1}|\psi_{2}\rangle|.
\end{equation}
Predictability is defined as the difference of the diagonal elements of the reduced density matrix $\rho^{\mathcal{A}}$,
\begin{equation}
    \mathcal{P}_{A} = |\rho^{A}_{11} - \rho^{A}_{22}| = |\langle\psi_{1}|\psi_{1}\rangle - \langle\psi_{2}|\psi_{2}\rangle|,
\end{equation}
which encodes the difference in probabilities of the outcomes for this two level system. For example, in case of two slit interference with single photon, it is the difference of probabilities of going through either slit. Using the above definitions of entanglement, coherence, and predictability one obtains,
\begin{align}
   C_{A|B}^{2} + \mathcal{C}_{A}^{2} + \mathcal{P}_{A}^{2} &= 4||\psi_{1}\rangle \wedge |\psi_{2}\rangle|^{2} + 4|\langle\psi_{1}|\psi_{2}\rangle|^{2} + |\langle\psi_{1}|\psi_{1}\rangle - \langle\psi_{2}|\psi_{2}\rangle|^{2} \\ 
    & = (\langle\psi_{1}|\psi_{1}\rangle + \langle\psi_{2}|\psi_{2}\rangle)^{2} = 1,
\end{align}
where, we have used the identity (\ref{LB}),
and the normalisation condition. Remarkably, the bipartite property concurrence and the single party property of coherence and predictability obeys a tight triality relation for a pure two party system,
\begin{equation}
    C_{A|B}^{2} + \mathcal{C}_{A}^{2} + \mathcal{P}_{A}^{2} = 1.
\end{equation}
For a separable pure state, concurrence is zero which results in the usual duality relation between the complementary characteristics of single party property of coherence and predictability \cite{tabish2015}. However, for a maximally entangled state concurrence is maximal and equal to one leading to zero coherence and predictability. Hence, it is observed that the single particle properties (coherence and predictability) are absent when the bipartite property of the system, entanglement is maximal which implies that the violation of Bell's inequality is intrinsically related with complementarity between the global and local properties \cite{jacob2010}. %It is worth noting that, if the reduced density matrix $\rho$ corresponds to the coherence-polarization matrix of the classical optics, the quantity $\mathcal{C}_{A}^{2} + \mathcal{P}_{A}^{2}$ is nothing but the degree of polarization, which is defined in terms of Stokes parameter as $P^{2} = \sum_{i =1}^{3} S_{i}^{2}$ where, $S_{i} = Tr(\sigma_{i}\rho)$. Hence, even in the classical optics scenario, where we do not have an analog of quantum entanglement, one can talk about the non-separability of various degrees of freedom, and define a measure of non-separability such as concurrence described above, and obtain an analogous complementary relation between the non-separability measure and the degree of polarisation \cite{eberly2016}.

It has been observed that for composite tri-partite pure states, the quantum correlation measure I-concurrence, completes the complementary relation between the local properties of the subsystem (such as, coherence and predictability) and the global entanglement of the composite system \cite{basso2020complete}. It would be interesting to further analyse the complementary characters of coherence and entanglement for the mixed states using the present approach. It would also be interesting to study the same for qutrit and higher dimensional systems, as it has strong implications in the multi-slit interferometry \cite{qureshi2021predictability}.    
 
\section{Conclusion} \label{sum}

In conclusion, the wedge product formalism of I-concurrence naturally leads to a geometric quantification of entanglement in terms of area spanned by the post-measurement vectors, wherein separability and entanglement is shown to be a consequence of vectors being parallel and non-parallel, respectively. Geometrical configurations corresponding to the separable and maximally entangled states is obtained by minimizing and maximizing the area spanned, and it was shown that the geometrical condition for maximal entanglement is orthogonality and equal norm of post-measurement state vectors corresponding to bi-partitions of the multiparty system. These conditions lead to the general form of the maximally entangled two qubit state, GHZ and GHZ like states. In the case of a three qubit system, the vanishing of concurrence corresponding to any two bi-partition is shown to be sufficient for tripartite separability. We derived the general conditions for maximally entangled states of two qudits, which lead to the generalised Bell states. Using the density matrix in terms of inner product between post-measurement vectors, a complementarity between global entanglement and local properties of the subsystems for pure two-qubit systems is discussed.  
%The present formalism can be extended to the intrapartite entanglement, where one deals with entanglement between various degrees of freedom for a single particle. However, the post-measurement vectors in such case need careful consideration, as they belong to different degrees of freedom of the same particle and independent unitary transformation needs careful study for the individual degrees of freedom. 

\begin{acknowledgements}
AKR and NKC acknowledge financial support of Inspire Scholarship provided by Department of Science
and Technology (DST), Govt. of India. SNS is thankful to UGC-CSIR for Junior Research Fellowship. PKP acknowledges support from DST, India through Grant No.: DST/ICPS/QuEST/Theme-1/2019/6.
\end{acknowledgements}

%\begin{acknowledgements}
%\end{acknowledgements}


\bibliographystyle{spmpsci_unsort}      % mathematics and physical sciences
\bibliography{sample}   % name your BibTeX data base

%\begin{thebibliography}{}
% etc
%\end{thebibliography}

\end{document}